\documentclass[11pt]{article}
\usepackage[margin=1in]{geometry}
\usepackage{amsmath,amssymb}
\usepackage{booktabs}
\usepackage{array}
\usepackage{graphicx}
\usepackage{hyperref}
\usepackage{natbib}
\usepackage{authblk}
\usepackage[utf8]{inputenc}
\usepackage[T1]{fontenc}
\usepackage{microtype}
\hypersetup{colorlinks=true, linkcolor=blue, citecolor=blue, urlcolor=blue}

\title{Who Gets Named: Citation Type Predicts Individual Naming by Grounded
Language Models, and a Roster Instrument Captures 0.5\% of It}

\author[1]{Dmitrij Żatuchin}
\affil[1]{Rankfor.AI OÜ, Tallinn, Estonia}

\date{July 2026}

\begin{document}
\maketitle

\begin{abstract}
Prior work on AI brand visibility measures the firm: does a model recommend a company, and
does that track its reputation. This study asks the question one level down, in categories
where the buyer picks a person. It issued 2{,}400 grounded API calls on
24 July 2026: 120 buyer-intent prompts, four models (GPT-5.6 Sol, Gemini 3.6 Flash, Perplexity
Sonar Pro, Grok 4.5), five iterations each, four European markets and five query languages.
Every response was coded for whether it named an individual professional, by a rule cascade
that never consults a roster and that drops detections resolving to a same-named American city
(precision 96.9\%; recall 61.7\% against roster ground truth, so every rate below is a lower
bound). All inference corrects for clustering within prompt: on the primary outcome the
intraclass correlation is 0.258, the design effect 5.90 and the effective $n$ 407 against a
nominal 2{,}400.

Models named an individual in 25.8\% of responses. Category dominates: real estate 35.4\% and
car dealerships 32.9\% against insurance 9.1\% ($\chi^2=159.3$, $p=5.8\times10^{-8}$ after the
design-effect correction). Models differ four-fold, from Grok 38.0\% to Gemini 9.3\%. Citation
type predicts naming and citation volume does not: naming responses cite the individual's own
site 2.6 points more often (95\% CI $+1.4$ to $+3.9$) and category portals 4.3 points more
often, cite social platforms less, and cite firm-owned pages at the same rate (44.1\% against
45.5\%). On nine matched translation pairs, English prompts named an individual in 36.7\% of
responses against 15.6\% for the same question in the local language (OR 3.14; clustered
$p=0.074$, so the direction is clear and the design cannot close it). A 939-person roster built
from public LinkedIn search matched 128 of
27{,}293 name-shaped mentions (0.47\%); 26 of the 939 people were ever named; and the
roster-derived rates, 0.0\% to 25.4\%, measure that overlap. Roster-based measurement of
individual AI visibility sees a small and unrepresentative slice of what models do.
\end{abstract}

\section{Introduction}
\label{sec:intro}

A generative model answering a buyer's question performs an act of gatekeeping: of everyone and
everything that could satisfy the query, it surfaces one name. Work on AI-mediated brand
visibility has established that this happens at the level of the firm, and that it diverges
from firm reputation in ways that matter commercially \citep{zatuchin2026reputationgap}. Whether
the same act operates at the level of the individual is unmeasured, and a large set of commercial
categories turns on it: the ones where the buyer's decision is ``which person.''

Real estate transactions, vehicle purchases and insurance placement share a feature: the buyer
picks a specific human being to act for them. A query such as ``best real estate agent in
Warsaw'' asks for a name. Whether a grounded model supplies one, and what decides that, is what
this paper measures.

\paragraph{The Recommendation Gap.} The construct is the distance between two quantities: how
often a generative model recommends an entity, and how far the entity's public standing
predicts that it should. Public standing is what independent sources record: review scores,
press coverage, professional registration, standing among peers. Recommendation is the share of
answers, across repeated draws from one prompt, that name the entity in reply to a buyer-intent
question. The gap is what remains of recommendation once standing is accounted for. Measuring
it takes three commitments, fixed before collection: a prompt set a buyer would plausibly type;
repeated sampling of every prompt, because one call to a non-deterministic model is one draw;
and a naming rule, fixed in advance, that decides what counts as a recommendation. Prior work
fixes the entity as a firm \citep{zatuchin2026reputationgap}. This paper fixes it as a person
and measures the recommendation term alone, holding standing constant by design: the question
is whether the model names anyone at all. This is a naming-rate study inside the Recommendation
Gap programme, and it computes no gap.

Two adjacent brand-level constructs sit nearby. The AIVO Paradox measures visibility against
purchase-recommendation win rate inside one model conversation \citep{derosen2026aivo}, and the
Linkage Gap measures what a model holds about a brand against what it deploys at the
recommendation turn \citep{sheals2026linkagegap}. Both compare two model behaviours. The
Recommendation Gap compares a model behaviour against an external record of standing.

Three research questions organise the study.

\begin{itemize}
    \item \textbf{RQ1 (naming rate).} In hyperlocal, person-centric buying categories, what
    share of grounded model responses name an individual professional, and what moves that
    share: category, market, model, or query language?
    \item \textbf{RQ2 (source predictors).} Does the composition of a grounded response's
    citations predict whether it names an individual?
    \item \textbf{RQ3 (cross-model agreement).} When more than one model names someone, do they
    agree on who?
\end{itemize}

A fourth question arrived from the data and is answered in
Section~\ref{sec:roster-instrument}: what does a roster-based instrument measure when it is
pointed at this problem, and how much of the answer does it see?

\section{Related work}
\label{sec:related}

\subsection{Brand-level AI visibility measurement}
\label{sec:related-brand}

A body of measurement work establishes that generative model recommendations diverge from
classical search rankings and from firm reputation. A twelve-brand study across five industries
found the correlation between narrative reputation quality and AI recommendation frequency
indistinguishable from zero (Spearman $\rho = 0.056$) \citep{zatuchin2026reputationgap}. A
fifty-brand, five-industry study using five-iteration replication (Section~\ref{sec:dice-roll})
found a mean Gini coefficient of 0.28 for recommendation-share concentration and 41.6\%
cross-model agreement on the top-recommended brand across three model families
\citep{zatuchin2026categoryownership}. A 128-brand, twelve-market, thirteen-language
citation-provenance study found that 85.7\% of the URLs a grounded model cites when discussing a
brand point to third-party domains the brand does not control \citep{zatuchin2026provenance}.

Larger surveys reach the same shape at scale: 100{,}000 prompt responses across 100 brands
\citep{kumar2026geoatscale}, 160{,}860 citation records across eight Chinese-language platforms
\citep{zhen2026chinesegse}, and 252{,}000 paired retrieval trials isolating which document
features win a citation \citep{vishwakarma2026whatgetscited}. An analysis of 2.8 million results
across 243 countries places AI search inside the wider information market
\citep{aral2026aisearch}, and separate work shows that incumbent brands hold an advantage in
what a model recommends \citep{chu2026incumbent}. Audit work on the citation layer itself finds
that source-cited answers support their claims less often than their presentation implies
\citep{venkit2025falsepromise}, and that Google's AI Overviews carry the same problem in a
consumer-health domain \citep{hu2025aioverviews}. Entity-level audit frameworks now exist for
bias in model outputs \citep{elbouanani2026entityaudit}. Throughout, the mechanism under study
is brand-level, and no named individual human is the object of recommendation.

\subsection{Individual-level naming in generative models}
\label{sec:related-naming}

Individual-level naming has been audited before. \citet{goethals2025fairness} score GPT-4,
Claude and Llama-3 on whether they name real Nobel laureates and leading figures in a field, and
report correctness, refusal and gender parity. \citet{barolo2025whosename} score six
open-weight models against American Physical Society and OpenAlex records;
\citet{espin2026scholarbench} extend that audit to 22 models under temperature,
prompt-constraint and web-search-grounded conditions; \citet{sanchez2026persona} vary persona
language and location across 43 models and six disciplines. All four ask who a model names when
asked for an expert, and all four ask it about scholars.
\citet{wang2026peoplesearchbench} verify the output of four commercial people-search products
against live web evidence, and \citet{baum2026electability} measures naming of political
candidates. \citet{sumbilon2026coldtest} tests one professional entity across six answer engines
and reports how much independent corroboration each engine needs before it names anyone.

Two literatures sit closer to a purchase. In industry, FlyDragon \citeyearpar{flydragon2026}
reports an average 8.4\% AI citation share for United States real estate agents across 12{,}400
responses, five answer surfaces and 192 metropolitan areas at five iterations per query. It
publishes neither its prompt set, its agent roster, its matching rule nor its model versions. A
trade column by one of its authors carries the result to practitioners
\citep{darani2026housingwire}. \citet{wilde2026miami} and companion studies
\citep{wilde2026pennsylvania,wilde2026newjersey} test 515 Miami-Dade and Broward businesses and
379 further health practices on local recommendation prompts and find 98.8\% to 99.6\% never
named, taking the business as the unit of naming. A vendor benchmark covers professional
services on 40 buyer questions and again scores firms \citep{authorityspecialist2026}, and a
simulation study evaluates legal-service discoverability without live model calls
\citep{clarity2025legalgeo}. \citet{sarkhedi2026protocol} published an open protocol for
auditing personal brand visibility in June 2026, and \citet{sarkhedi2026avf} a six-signal
framework for it, neither with a primary measurement attached; a companion practitioner
framework reports 523 consultancy engagements and queries no model \citep{sarkhedi2026pbei}.
\citet{singh2026invisible} follows one individual through seven months of entity
disambiguation. \citet{carr2026prereg} registered a five-model study of local business
recommendation in February 2026 that had posted no data, code or materials when this paper was
written.

This study takes the question to the individual inside commercial buyer-intent categories, where
the named person is the object of the purchase decision, and outside the United States, across
four European markets and five query languages. To the author's knowledge it is the first
measurement of individual-level naming by grounded generative models in commercial buyer-intent
categories that publishes its prompt set, its roster construction rule, its detection rule and
its matching procedure, and the first conducted outside the United States. The disclosure
qualifier carries the weight: FlyDragon holds the ground on individuals in a commercial category
and withholds the method.

\subsection{Individual-level algorithmic visibility}
\label{sec:related-individual}

A distinct literature examines algorithmic visibility of individual workers on digital
platforms as a labour-market outcome. Algorithmic management research treats platform-mediated
visibility, task allocation and ranking as core mechanisms by which platform workers are
governed, a body of work large enough to have supported a 103-source interdisciplinary
systematic review \citep{chigbu2026algorithmic}. Within it, platform-mediated visibility is a
site of demographic disparity. A study of 108 online freelancers found systematic gender- and
race-based patterns in how digital labour platforms make individual workers visible to clients
\citep{munoz2023platformization}. A large observational study of TaskRabbit and Fiverr put the
disparity at an incidence rate ratio near 0.90 on review counts \citep{hannak2017bias}.
Generative systems carry demographic structure of their own: a cross-model audit of 41
occupations finds stable race and gender patterns in the personas models produce
\citep{vanderlinden2025modalworker}.

That literature has been extended to generative answer surfaces for scholars and public figures
\citep{goethals2025fairness,barolo2025whosename,espin2026scholarbench,sanchez2026persona} and to
local businesses as entities \citep{wilde2026miami}. It has not been extended to commercial
service professionals as individuals, where the visibility act sits inside a purchase decision.
Section~\ref{sec:limitations} states what this design can and cannot say about the demographic
dimension.

\subsection{The dice-roll replication standard}
\label{sec:dice-roll}

Because a single call to a non-deterministic generative model is a single draw from a
distribution, this study adopts the replication design established and power-analysed across
five prior studies, roughly 190{,}000 observations \citep{zatuchin2026diceroll}. That work shows
$n=5$ replications detects only large effects ($d>1.2$, power $>0.80$), with metric convergence
reaching 80\% of its asymptotic value by $n=7$ and test-retest reliability crossing 0.70 at
$n\geq8$. Independent work reaches the same conclusion from other angles: prompt paraphrase
alone moves measured visibility enough to need a confidence-aware protocol
\citep{kirichenko2026reproducible}, prompt wording shifts brand visibility across 288
human-written prompts \citep{ehrlinspiel2026prompttracking}, and identical client profiles draw
different professional advice on repeat asking \citep{agliata2026advisor}. This study uses the
$n=5$ tier throughout. Section~\ref{sec:stats} shows that clustering within prompt bounds
precision here.

\subsection{Language and morphology in grounded output}
\label{sec:related-language}

A line of work measures cross-language divergence in grounded brand recommendations directly. A
sixty-six-brand study spanning twelve European languages (35{,}640 responses, three models)
found that AI-constructed brand reputation varies by query language, and that switching from
English to a brand's own language raises recommendation frequency far more for local brands
(+0.80) than for global corporations (+0.15) \citep{zatuchin2026langblindspot}. A
variance-components decomposition (12{,}933 responses, 20 brands, 8 languages, 3 models) found
query language alone explains 26.5\% of response variance against brand identity's 1.5\%
\citep{zatuchin2026variance}. Section~\ref{sec:language} extends that work to whether an
individual gets named at all.

Polish and Lithuanian inflect personal surnames by sentence role, which puts entity boundaries
out of step with token boundaries and makes the matching task need morpheme-level treatment
\citep{bareket2020nemo}. No prior work in this literature reports a name-matching procedure
designed for surnames that inflect by grammatical case. Section~\ref{sec:method-matching}
reports the check this study ran and states its limit; \citet{christen2006names} benchmarks the
personal-name matching techniques a general solution would use.

\section{Method}
\label{sec:method}

\subsection{Design and prompts}
\label{sec:method-prompts}

The design crosses four markets (Poland, Ireland, the Netherlands, and a combined
Lithuania/Estonia ``Baltics'' market, treated jointly because each economy is individually small
and because prior work in this programme already treats the Baltic states as one unit
\citep{zatuchin2026reputationgap}) with three categories: real estate agents, car dealership
sales representatives and insurance brokers.

For each (market, category) pair the study used seven English buyer-intent prompts: five short
generic variants (``Who's the best [category] in [city]?'', ``Recommend a [category] in
[city]'') and two attribute-carrying variants, an expat-buyer framing and a first-time-buyer
framing. That split matches this programme's own finding on the length and attribute-carrying
share of real buyer prompts \citep{zatuchin2026reputationgap}. Poland, the Netherlands and each
Baltic city also received three local-language prompts per category: two short generic
variants and one attribute-carrying variant, in Polish, Dutch, Lithuanian and Estonian. Ireland
received English prompts only. This gives 120 unique prompts: Poland 30, Ireland 21, Netherlands
30, Baltics 39. Appendix~\ref{sec:prompts} prints all 120 verbatim.

Nine English prompt slots have an exact local-language translation in the same city and
category, and the rest do not. Section~\ref{sec:language} treats those nine as the inferential
language test and the full set as descriptive.

\subsection{Models and collection}
\label{sec:method-models}

Four providers span the commercially deployed grounded-model families as of July 2026: GPT-5.6
Sol (OpenAI), Gemini 3.6 Flash (Google), Perplexity Sonar Pro, and Grok 4.5 (xAI). Each model
identifier and grounding mechanism was live-verified against the provider's production API
immediately before collection, with a single test call confirming a grounded response carrying
citation metadata.

\textbf{A fix made mid-collection.} The first 600 OpenAI calls used the Chat Completions
endpoint. Chat Completions does not support genuine web search for the GPT-5.x family: the
\texttt{web\_search\_options} parameter is restricted to a separate family of dedicated
\texttt{*-search-preview} models, so those calls were an ungrounded, training-knowledge-only
pass, inconsistent with the other three providers. The implementation was corrected to the
Responses API with a genuine \texttt{web\_search} tool invocation, and all 600 OpenAI calls were
re-collected before any analysis reported here. The correction is disclosed because the failure
mode is generic and cheap to repeat: an API endpoint silently not supporting the grounding tool
a researcher assumes it does.

Each (prompt, model) pair was queried five times, giving $120 \times 4 \times 5 = 2{,}400$
calls. All 2{,}400 were issued in one continuous session on 24 July 2026.

Every call requested live web-search grounding. Grounding did not always arrive. Of the 2{,}400
responses, 2{,}223 (92.6\%) carry at least one citation and 177 (7.38\%) carry none. The
shortfall is concentrated. 137 of the 177 are OpenAI, which returned no citation on 22.83\% of
its 600 calls (95\% CI 19.65 to 26.36), against Gemini 31, Perplexity 7 and Grok 2.
Zero-citation responses are also uneven by market: Ireland 78 of 420 (18.6\%), Poland 52 of 600,
Netherlands 29 of 600, Baltics 18 of 780. Every market-level comparison below therefore carries
a differential-grounding component, and Section~\ref{sec:stats} reports a grounded-only
sensitivity run.

A further 140 responses, all OpenAI, ask a disambiguating question in place of an answer
(Ireland 66, Poland 43, Netherlands 31; operational rule: under 700 characters and containing a
question mark). They ask which Dublin, Warsaw or Amsterdam the user means, which is itself a
result and is reported in Section~\ref{sec:geo}. Four clarifying responses also carry citations,
so the two exclusions overlap by four and the analysis population excluding both is 2{,}219
(92.5\%).

\subsection{The professional roster}
\label{sec:method-roster}

The study also built a roster of 939 named individuals across the same four markets and three
categories, by LinkedIn people search in Classic mode using keyword and location facets. A Sales
Navigator seat, which would have allowed further faceted filtering, was not available.
Table~\ref{tab:roster} gives the per-cell sizes.

\begin{table}[h]
\centering
\caption{Roster composition by market and category.}
\label{tab:roster}
\begin{tabular}{lccc}
\toprule
Market & Real estate & Car dealers & Insurance \\
\midrule
Poland & 98 & 79 & 78 \\
Ireland & 73 & 75 & 71 \\
Netherlands & 89 & 75 & 75 \\
Baltics (LT+EE) & 80 & 79 & 67 \\
\bottomrule
\end{tabular}
\end{table}

\textbf{Inclusion rule.} A candidate entered the roster if and only if, at collection time, a
citable, resolving \texttt{linkedin.com/in/} profile URL named the individual in the target
role. The rule is deliberately narrow, for two reasons independent of statistical convenience.
It keeps every roster row traceable to one external URL, so a removal request is a one-line
action. It also keeps the sample inside the most defensible category of public professional
data under the Art.~6(1)(f) legitimate-interest basis: data the subject or their employer
published so that a prospective client would find it. No purchased contact records, no
enrichment-tool outputs without a resolving public profile, and no source that is not itself a
citable public URL entered the file. This is a convenience sample bounded by LinkedIn's own
search surfacing, and Section~\ref{sec:roster-instrument} reports what that turns out to mean.

\textbf{What the roster is, measured.} The roster carries no city field. Its \texttt{location}
column holds the subject's self-declared LinkedIn region, and for 393 of 939 rows (41.9\%) that
string is the bare market name; 497 rows (52.9\%) resolve below country level. Counting a row as
city-bounded when the prompted city appears in \texttt{location} or \texttt{headline}, and
allowing the Polish stem \emph{warszaw} and the Lithuanian genitive \emph{Vilniaus}, gives
Poland 15.7\%, Ireland 16.4\%, Netherlands 23.4\% and the Baltics 35.0\%. Ireland's figure
cannot be read: its one named row carries the market-name fallback, so Ireland supplies no test
of city-boundedness.

Feeding every roster name through the extractor inside a neutral carrier sentence, 759 names
(80.8\%) recover exactly, 46 (4.9\%) only under the fuzzy fallback, and 134 (14.3\%, 95\% CI
12.2 to 16.6) never recover. The rate is not uniform: Poland 5.1\%, Ireland 14.6\%, Baltics
4.0\%, Netherlands 33.5\%. The dominant cause is the Dutch \emph{tussenvoegsel}: 67 of 239 Dutch
rows carry a lower-case particle and all 67 are unrecoverable. The rest are names above three
tokens (29), punctuation or emoji in the name field (24), and non-ASCII initial capitals (7).
The 14.3\% is a floor, measured in the friendliest carrier position.

Thirty rows (3.2\%) trade under a firm name built from the individual's own name. For those the
person-versus-firm distinction is undecidable from the output string alone, and the bias runs
toward under-counting: the firm string carries extra capitalised tokens (``\emph{Surname}
Makelaardij''), which the extractor emits as a three-word span matching no roster row. Three of
the 26 individuals a model ever named hold such a firm.

\subsection{Two ways of deciding that a response named someone}
\label{sec:method-matching}

\textbf{Roster matching (the instrument).} Model output was scanned for name-shaped spans and
matched against the 939-person roster by exact normalised string match, with a fuzzy fallback
(surname edit distance $\leq 1$, given-name prefix match). This produced 128 matched mentions
across 114 of the 2{,}400 responses. Fifteen of the 128 are fuzzy matches, not exact ones, and ten
responses rest entirely on a fuzzy match; restricting to exact matches moves the Baltics rate
from 13.97\% to 12.69\% (99/780) and leaves the other three markets unchanged, so nothing
reported here depends on the fuzzy rule.

Because Polish and Lithuanian inflect surnames by sentence role, a stemmed re-check ran on the
Polish subsample before any market difference was treated as real. Eight hundred and seventy
name-shaped strings surviving a company-and-generic-phrase filter, drawn from 2{,}779 distinct
unmatched strings and 6{,}976 raw mentions, were reduced to five-character stems on both given
name and surname and compared against a stemmed Poland roster. This produced zero additional
matches. A positive control confirmed the pipeline recovers repeated identical matches in the
Baltics subsample. The check is reported because it ran, and its limit is stated with it: it
tested the 870 strings a company-and-generic filter had already passed, and that filter is the
kind of instrument whose coverage is in question. Five-character truncation was chosen because
it over-matches, which is what a test for a false zero needs; a general solution should use the
phonetic and similarity-based methods benchmarked by \citet{christen2006names}.

A second defect is measured. The shipped matcher used an ASCII word class, so
any name beginning with a non-ASCII capital could never match. Scanning every response for the
literal roster names of its own market and category, diacritic-folded and word-bounded, finds 43
occurrences across six people that the matcher never recorded. One Baltic car-sales professional
is named unambiguously as a person in 19 separate responses and scores zero every time.
Correcting it moves Baltics real estate from 25.38\% to 25.77\%, Baltics car dealers from
16.15\% to 19.62\%, and Netherlands car dealers from 0.00\% to 1.00\%. Counting the
eponymous-firm strings as person namings instead takes Netherlands real estate to 10.5\% and the
Dutch market total to 3.8\%, which is the honest upper bound on the same cell.

\textbf{Roster-free detection (the primary outcome).} The roster matched 128 of 27{,}293
name-shaped mentions the models emitted, 0.47\%, and 128 of 1{,}876 spans the detector resolves
to individuals, 6.8\%. Any outcome defined by roster membership therefore measures the
intersection of two populations. The primary outcome is a rule cascade applied to the response
text, which never opens the roster. Six stages run in order: markdown stripping; candidate
generation, meaning two or three capitalised tokens plus nobiliary particles; a shape gate; a
lexical gate against corporate, geographic and generic blocklists in five languages plus street
suffixes; a company-adjacency test on the whole enclosing capitalised run; and a
positive-evidence requirement. That last stage demands a given-name gazetteer hit, a role
apposition, or an address verb or title before the span. An eponym rule then removes spans that
are firm names shaped like personal names, tested against company-shaped runs in the same
response and market and against brand keys derived from cited domains. A final stage drops
detections resolving to a business or person in a same-named American city: Dublin, Ohio;
Warsaw, Indiana; Amsterdam, New York. Each such domain was adjudicated against the live site,
the stated address and, where the site was unclear, a company register. That stage defines the
geo-corrected outcome used throughout.
Appendix~\ref{sec:pseudocode} states the cascade as pseudocode.

Two measured properties bound everything that follows. \emph{Precision} is 96.9\% (155 of 160
detections audited in two independent passes, 95\% CI 92.9 to 98.7). \emph{Recall} against the
128 roster mentions is 61.7\% (Baltics 62.6\%); the misses are gazetteer gaps on given names
such as \emph{Xander}, \emph{Raimond} and \emph{Leho}. Every naming rate below is a lower bound,
and the gap is larger in the Baltics than elsewhere, which works against the market comparison
in Section~\ref{sec:rq1}. A corpus-induced gazetteer was tested as a recall fix
and rejected: it added 86 names, all spurious, and inflated every market.

\subsection{Citation source classification}
\label{sec:method-citations}

Every citation URL was reduced to a registrable hostname, giving 20{,}580 citations across
2{,}182 distinct domains. The classification reported here replaces the one applied at
collection time, and both are described, because the difference changes the answer to RQ2.

\textbf{What the collection pipeline did.} The collection-time classifier tested each hostname
against an allowlist of roughly forty domains and one registry-hint pattern, and returned
\texttt{employer-owned page} as its else-default. Against 2{,}182 observed domains that default
absorbed 2{,}110 of them and 17{,}945 of 20{,}580 citations (87.2\%), including
\texttt{reddit.com}, the single most-cited domain in the corpus at 746 citations. Three
substring bugs compound it: the registry hint \texttt{fi.ee} matches \texttt{delfi.ee} (4
citations), the bare hint \texttt{register} matches \texttt{inforegister.ee} (35), and the
social-domain entry \texttt{x.com} matches \texttt{carfax.com} (21), \texttt{yandex.com} (10)
and \texttt{global.remax.com} (2). The failure mode deserves a name: an allowlist with a
substantive else-default reports its own default category as a finding.

\textbf{The replacement.} All 2{,}182 domains were classified under a twenty-category taxonomy
set out in a written coding manual, with a controllability axis (owned, paid, earned,
not-acquirable) recorded alongside. The 800 highest-volume domains, carrying 18{,}201 citations
(88.4\% of the corpus), were classified one by one against the live site; the remaining 1{,}382
were assigned by published rule and flagged low-confidence. The taxonomy separates three things
the old \texttt{employer-owned page} bucket conflated: the firm's own site (\texttt{firm-site}),
the individual professional's own site (\texttt{individual-site}), and category portals
(\texttt{vertical-portal}). Four hundred and seventy-eight domains sit in
\texttt{unclassified}; nothing defaults into a substantive category. The classification is
LLM-assisted and is described as such in the deposit.

\textbf{Adversarial audit.} An independent pass re-read all 100 top-ranked domains against the
cited paths and five live pages, and found no third-party platform coded as a firm's own page. A
portal-signature test across all 472 \texttt{firm-site} and \texttt{individual-site} rows in
ranks 1 to 800 found no host coded as a firm showing platform behaviour. Eleven rows in ranks 1
to 800 carry a label the audit would change, together 179 citations, 0.87\% of the corpus. In
the rule-coded tail, one rule labelled 359 hosts as a firm's own site because a trade word
appears in the domain string, carrying 650 citations. That reproduces the original defect in
miniature. Demoting every such row to \texttt{unclassified} leaves all four surviving contrasts
in Section~\ref{sec:rq2} unchanged to two decimal places, and moves only \texttt{firm-site},
a null under both codings.

\textbf{The individual-site boundary.} A later pass withdrew the inference that a person's name
in a domain, or a sole-trader registration, implies that one person delivers the service. A
business named after someone can employ staff. Under a corrected rule requiring positive
evidence of solo operation, verified against each live site and its sitemap, 41 of 70
\texttt{individual-site} domains left the class, which fell from 70 domains and 706 citations to
29 and 319. The contrast resting on that boundary lost about one percentage point and kept its
sign in every treatment of the unresolved rows (Section~\ref{sec:rq2}).

\textbf{Employer identity.} Testing whether naming responses cite the named person's own
employer requires resolving each named individual to an employer domain, which the roster
headline supports for 313 of 939 individuals (33.3\%; Ireland 53.0\%, Baltics 51.3\%,
Netherlands 30.5\%, Poland 3.1\%). On the 26 individuals ever named, audited one by one,
citations landing on that person's own employer domain number 34 of 975 (3.49\%, 95\% CI 2.10 to
5.13), and 25 of 114 naming responses (21.9\%) cite it at least once. For 51.1\% of
person-citation pairs the roster row names no employer at all, so the claim is undefined for the
majority of them.

\subsection{Statistical approach}
\label{sec:stats}

Each of the 120 prompts was answered 20 times (four models, five iterations), so responses are
not independent. On the primary geo-corrected outcome the intraclass correlation within prompt
is 0.258, which gives a design effect of $1 + (\bar{m}-1)\rho = 5.90$ at $\bar{m}=20$ and an
effective sample size of 407 against a nominal 2{,}400. On the roster-matched outcome the same
three quantities are 0.266, 6.06 and 396. Clustering is what bounds precision in this design. The nominal $n$
overstates it more than fivefold.

Single proportions carry Wilson intervals, and zero cells carry Clopper-Pearson bounds.
Between-prompt comparisons are widened by the design effect before testing (Rao-Scott), with the
design effect computed within arm and size-weighted. Pooling the two arms is wrong here: every
prompt sits entirely inside one arm of every between-prompt contrast, so a pooled design effect
absorbs the arm mean difference into between-prompt variance and penalises large effects
hardest. The two estimators differ materially: on the Vilnius language contrast they give 3.83
within arm against 5.35 pooled.

Model is the exception. Because model is crossed with prompt, prompt clustering acts as blocking
for model contrasts and improves their precision, so a between-cluster design effect would
correct in the wrong direction. Model contrasts are tested by exact within-prompt randomisation
(20{,}000 permutations of the outcome inside each prompt, model labels fixed) and by
Cochran-Mantel-Haenszel stratified on prompt, and the two agree.

Families of comparisons are corrected by Holm. Citation-composition intervals come from a
cluster bootstrap that resamples whole prompts, 10{,}000 draws, seed 20260726. A bootstrap
over responses understates those intervals by roughly a third. Where an arm
contains zero events (Poland under the roster instrument, 0 of 600) the report gives the
Clopper-Pearson upper bound and the rule of three, because both GEE and unpenalised logistic
regression return separation artifacts on that cell.

The multivariable model is a logistic mixed model with a random intercept on prompt, fitted by
variational Bayes, with market, category, model and query language as fixed effects
(Table~\ref{tab:glmm}). A GEE with an exchangeable working correlation clustered on prompt is
fitted alongside as a check on the standard errors.

Five sensitivity analyses run alongside the primary results. They restrict to the 2{,}223
responses returning at least one citation; to the 2{,}219-response analysis population, which
also drops the 140 clarifying responses; to exact-match-only roster matching; to the detector
without its geographic stage; and to the citation taxonomy with the tail rule demoted. None
reverses a reported direction.

\section{Results}
\label{sec:results}

\subsection{RQ1: category dominates, market mostly does not separate}
\label{sec:rq1}

\begin{figure}[htbp]
\centering
\includegraphics[width=\textwidth]{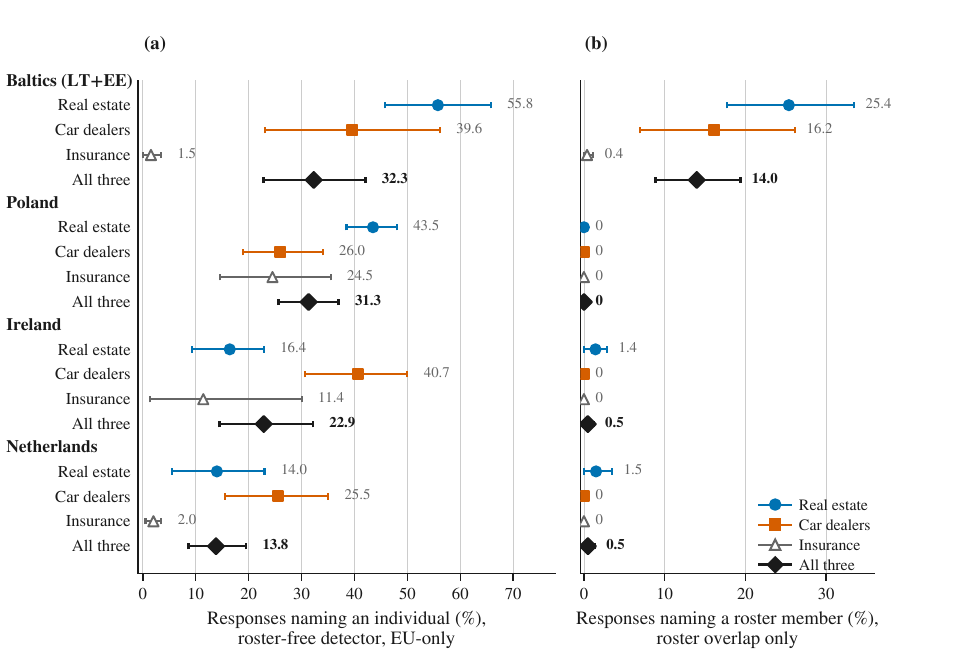}
\caption{Naming rate by market and category. Panel (a) plots the primary outcome, the
geo-corrected roster-free detector, and matches Table~\ref{tab:naming}. Panel (b) plots the
roster-matched diagnostic on its own scale. Intervals are prompt-level cluster bootstrap,
10{,}000 draws.}
\label{fig:naming}
\end{figure}

Models named an individual professional in 619 of 2{,}400 responses, 25.79\% (95\% CI 24.08 to
27.58). Table~\ref{tab:naming} gives the rate by market and category, and
Figure~\ref{fig:naming} plots the same cells with prompt-level cluster-bootstrap intervals.

\begin{table}[h]
\centering
\caption{Naming rate by market and category, primary outcome (geo-corrected, roster-free
detector). Cells hold 200 responses in Poland and the Netherlands, 140 in Ireland, 260 in the
Baltics. Parenthesised values give the rate before the geographic stage, where it differs.}
\label{tab:naming}
\begin{tabular}{lcccc}
\toprule
Market & Real estate & Car dealers & Insurance & All \\
\midrule
Poland      & 43.5\% (44.0) & 26.0\% (28.0) & 24.5\% (25.5) & 31.3\% (32.5) \\
Ireland     & 16.4\% (29.3) & 40.7\% (44.3) & 11.4\%        & 22.9\% (28.3) \\
Netherlands & 14.0\%        & 25.5\% (28.0) & 2.0\%         & 13.8\% (14.7) \\
Baltics     & 55.8\%        & 39.6\%        & 1.5\%         & 32.3\% \\
\midrule
All         & 35.4\% (37.8) & 32.9\% (34.6) & 9.1\% (9.4)   & 25.8\% (27.3) \\
\bottomrule
\end{tabular}
\end{table}

\textbf{Category.} Real estate (35.4\%) and car dealerships (32.9\%) are indistinguishable from
each other ($p = 0.62$) and both far exceed insurance (9.1\%). Real estate against insurance
gives $\chi^2 = 159.3$ and $p = 5.8\times10^{-8}$ at a within-arm design effect of 5.41; car
dealers against insurance gives $p = 4.9\times10^{-7}$ at 5.37. Before the geographic stage the
same two contrasts give $\chi^2 = 178.8$, $p = 4.2\times10^{-9}$ and $p = 1.7\times10^{-7}$.
This is the largest and best identified effect in the study. It holds inside every market, with
one qualification: insurance sits near floor in the Baltics (1.5\%) and the Netherlands (2.0\%)
and stays meaningfully positive in Poland (24.5\%) and Ireland (11.4\%). The Irish insurance
figure rests on two recurring named experts, one appearing in 15 of 140 Irish insurance
responses and one in 5. Two people carry that figure, so it is a statement about them.

\textbf{Market.} Poland 31.3\% (95\% CI 27.8 to 35.2), Baltics 32.3\% (29.1 to 35.7), Ireland
22.9\% (19.1 to 27.1), Netherlands 13.8\% (11.3 to 16.8). Pairwise at within-arm design effects
with Holm over six comparisons, two contrasts clear: Poland against the Netherlands
($p_{\text{Holm}} = 0.0003$) and the Baltics against the Netherlands
($p_{\text{Holm}} = 0.012$). Ireland against the Netherlands does not
($p_{\text{Holm}} = 0.31$), and none of the Poland, Baltics and Ireland contrasts come close
(all $p_{\text{Holm}} \geq 0.31$). The defensible statement is that the Netherlands is last and
the other three do not separate in this design. Design effects differ sharply by market (Poland
2.31, Ireland 4.81, Netherlands 4.16, Baltics 8.86), which is informative on its own: Baltic
responses agree with each other inside a prompt far more than Polish ones do.

\subsection{RQ1b: models differ more than markets}
\label{sec:models}

\begin{figure}[htbp]
\centering
\includegraphics[width=0.85\textwidth]{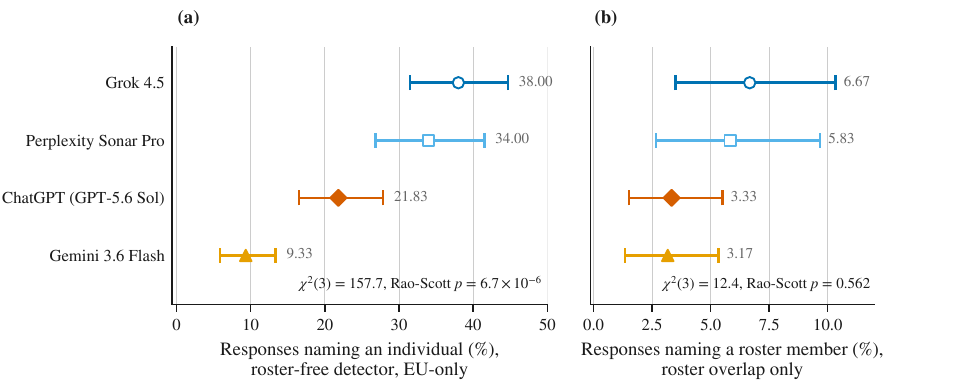}
\caption{Naming rate by model under both outcome definitions.}
\label{fig:bymodel}
\end{figure}

Grok names an individual in 38.0\% of its 600 responses, Perplexity 34.0\%, GPT-5.6 Sol 21.8\%
and Gemini 9.3\% (Table~\ref{tab:models}, Figure~\ref{fig:bymodel}). Because every one of the
120 prompts was answered by all four models, model is a within-prompt factor, and the test is
exact within-prompt randomisation: $\chi^2 = 157.7$, permutation $p < 0.0001$ over 20{,}000
draws. Cochran-Mantel-Haenszel stratified on prompt agrees on every pair except Grok against
Perplexity: Grok against Gemini $z = 13.48$, Perplexity against Gemini $z = 12.30$, GPT against
Gemini $z = 6.85$ ($p = 7.2\times10^{-12}$), Grok against GPT $z = 7.49$, Perplexity against GPT
$z = 5.79$, and Grok against Perplexity $z = 1.80$, $p = 0.071$.

\begin{table}[h]
\centering
\caption{Naming rate by model, $n=600$ per model, Wilson 95\% intervals. The roster-matched
column is the diagnostic of Section~\ref{sec:roster-instrument}.}
\label{tab:models}
\begin{tabular}{lccc}
\toprule
Model & Primary outcome & 95\% CI & Roster-matched \\
\midrule
Grok 4.5             & 38.0\% & 34.2 to 42.0 & 6.7\% \\
Perplexity Sonar Pro & 34.0\% & 30.3 to 37.9 & 5.8\% \\
GPT-5.6 Sol          & 21.8\% & 18.7 to 25.3 & 3.3\% \\
Gemini 3.6 Flash     &  9.3\% &  7.3 to 11.9 & 3.2\% \\
\bottomrule
\end{tabular}
\end{table}

The GPT-against-Gemini contrast deserves its own line, because it is the one model comparison
this design was already powered for: 21.83\% against 9.33\%, a gap of 12.5 points, OR 2.71,
Fisher $p = 2.6\times10^{-9}$, Cohen's $h = 0.351$, needing 359 responses per arm at 80\% power
against 600 held.

The same comparison on the roster-matched outcome is 3.33\% against 3.17\%, a gap of 0.17
points, Fisher $p = 1$, and the omnibus model effect there falls from $\chi^2 = 12.4$ to a
Rao-Scott $\chi^2 = 2.05$, $p = 0.562$. The roster-matched measure compresses all four models
toward zero and cannot separate them. Only the roster-free ordering is reported as a result.

\subsection{RQ2: citation type predicts naming, citation volume does not}
\label{sec:rq2}

Responses that named an individual and responses that did not cite a similar number of sources.
Under the roster instrument the means are 8.55 and 8.58, with a cluster-bootstrap interval on
the difference of $-1.34$ to $+1.17$ and $d = -0.003$; under the primary outcome they are 7.90
and 8.81. Volume carries no signal.

\begin{figure}[htbp]
\centering
\includegraphics[width=\textwidth]{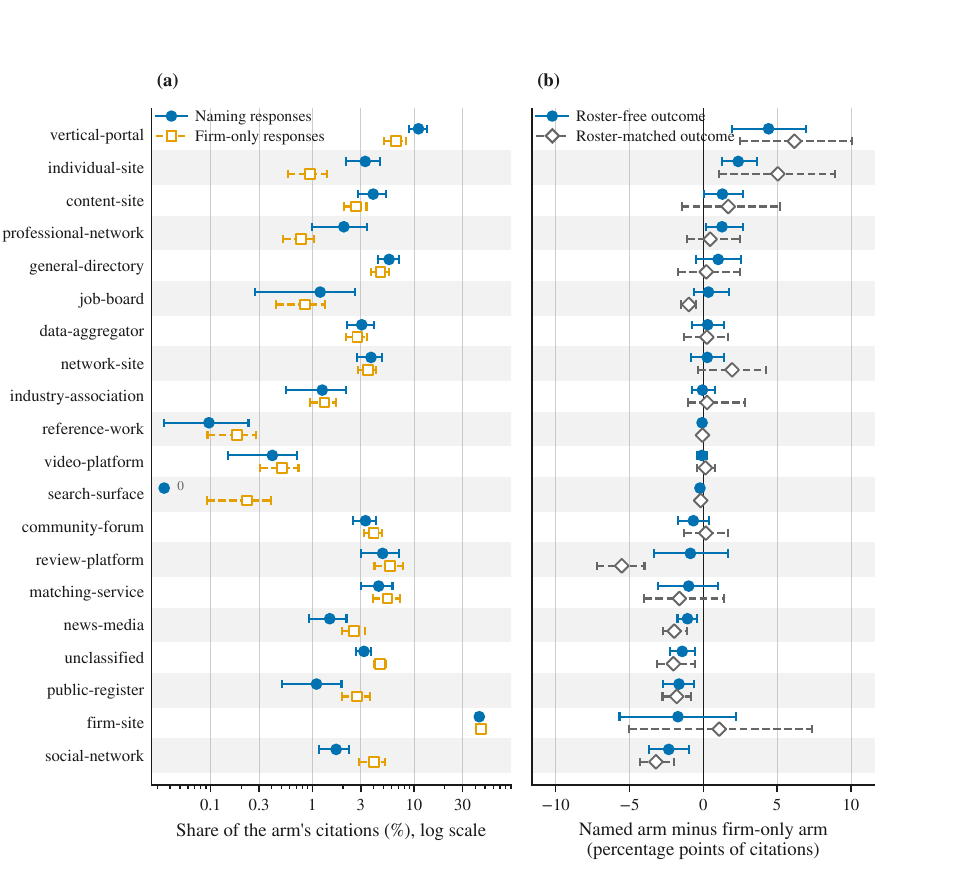}
\caption{Citation composition by source type, naming against firm-only responses, all twenty
categories. The roster-free series here is the detector before its geographic stage, where the
individual-site gap is $+2.36$ and the vertical-portal gap $+4.42$; Table~\ref{tab:citations}
gives the geo-corrected values, $+2.58$ and $+4.27$, which is the difference between the two.}
\label{fig:citations}
\end{figure}

Composition does. Table~\ref{tab:citations} gives the gap in each source type's share of an
arm's citations, under both outcome definitions, with prompt-level cluster-bootstrap intervals.
Figure~\ref{fig:citations} plots the full twenty-category breakdown.

\begin{table}[h]
\centering
\caption{Citation composition: gap in percentage points between naming and firm-only responses
(named minus firm-only). Intervals are prompt-level cluster bootstrap, 10{,}000 draws, seed
20260726. Bold marks the three contrasts this section reports as findings. Six rows exclude zero
under both definitions; the three in bold are the ones that also hold under the geographic
correction and the tail-rule sensitivity and carry a mechanism the text can name.}
\label{tab:citations}
\begin{tabular}{lcc}
\toprule
Source type & Roster instrument & Primary outcome \\
\midrule
\textbf{individual-site}  & \textbf{$+5.05$ [$+1.08$, $+8.90$]}  & \textbf{$+2.58$ [$+1.41$, $+3.91$]} \\
\textbf{vertical-portal}  & \textbf{$+6.17$ [$+2.49$, $+10.08$]} & \textbf{$+4.27$ [$+1.73$, $+6.91$]} \\
\textbf{social-network}   & \textbf{$-3.21$ [$-4.28$, $-2.01$]}  & \textbf{$-2.19$ [$-3.50$, $-0.85$]} \\
review-platform           & $-5.53$ [$-7.23$, $-3.98$]           & $-1.70$ [$-3.96$, $+0.61$] \\
public-register           & $-1.80$ [$-2.77$, $-0.86$]           & $-1.56$ [$-2.63$, $-0.52$] \\
news-media                & $-1.97$ [$-2.73$, $-1.08$]           & $-1.04$ [$-1.73$, $-0.35$] \\
firm-site                 & $+1.08$ [$-5.01$, $+7.34$]           & $-1.41$ [$-5.45$, $+2.62$] \\
general-directory         & $+0.20$ [$-1.71$, $+2.51$]           & $+1.36$ [$-0.15$, $+2.93$] \\
unclassified              & $-2.03$ [$-3.14$, $-0.59$]           & $-1.50$ [$-2.32$, $-0.65$] \\
\bottomrule
\end{tabular}
\end{table}

\textbf{Firm-owned pages carry no signal.} Naming responses take 44.05\% of their citations from
firm sites and firm-only responses 45.46\%, and the gap interval spans zero; under the
classification used before the individual-site re-adjudication the same two shares are 44.3\%
and 44.1\%. The original analysis of this corpus reported 95.5\% against 86.8\% and read the
difference as the paper's mechanism. That split was an artifact of the else-default described in
Section~\ref{sec:method-citations}, which swept 87.2\% of citations into a bucket labelled
\texttt{employer-owned page}. On the 26 individuals ever named, citations landing on that
person's own employer domain number 34 of 975, 3.49\%.

\textbf{What does distinguish them} is the individual's own site and the category portal,
together with the absence of social platforms. Under the primary outcome, naming responses take
3.52\% of their citations from a site one person delivers, against 0.94\% for firm-only
responses, and 11.02\% from vertical portals against 6.75\%. Both survive the change of outcome
definition, the geographic correction and the tail-rule sensitivity. Without the geographic
stage the same two gaps are $+2.36$ ($+1.26$ to $+3.64$) and $+4.42$ ($+1.93$ to $+6.98$). The individual-site
contrast survived a harder test as well: withdrawing the inference that a person-named domain
implies a solo operator removed 41 of 70 domains from the class and cost the contrast about one
point, and it still excludes zero in all three treatments of the unresolved rows.

Two caveats belong in the same breath. First, 108 of the 114 naming responses under the roster
instrument sit in two cells, Baltics real estate and Baltics car dealers, so every pooled
contrast on that outcome is partly a market and category comparison. Stratified inside that
stratum the individual-site gap falls to $+3.91$ ($-1.51$ to $+9.40$) and vertical-portal to
$+1.86$ ($-1.59$ to $+5.64$), whose intervals at $n=108$ are too wide to establish an
absence. Second, the roster-matched individual-site contrast rests on 62 named-arm citations
across nine domains and eight prompt clusters, every one of them Baltics real estate, and one
domain carries 27 of the 62. The primary outcome does not have that problem to the same degree,
with 619 naming responses spread across all four markets, and it is where the finding should be
argued.

\textbf{Review platforms do not survive.} The pooled roster-matched gap of $-5.53$ looks like
the strongest row in the table, and it collapses under both stress tests: $-0.82$ ($-1.67$ to
$-0.03$) inside the Baltic stratum, and $-1.70$ ($-3.96$ to $+0.61$) under the primary outcome.
It is close to pure market and category composition, and it is not reported as a finding. Social
platforms are the one contrast that strengthens under stratification ($-5.83$, $-7.76$ to
$-3.39$) and holds under the change of outcome.

\subsection{Query language: a within-city effect and an informative null}
\label{sec:language}

Nine English prompt slots have an exact local-language translation in the same city and
category. Those nine pairs are the inferential test, because they hold market, category, city,
model, iteration count and question wording fixed by construction, and vary only the language of
the query.

\textbf{On the matched pairs, and on the primary outcome, English named an individual in 66 of
180 responses (36.67\%) against 28 of 180 (15.56\%) for the translation.} Fisher
$p = 7.3\times10^{-6}$, OR 3.14. The direction is the same in all three pair groups: 40.0\%
against 13.3\% in Vilnius, and 35.0\% against 16.7\% in each of the two Tallinn groups.

That Fisher test treats 360 responses as independent, and they are not. Under the design-effect
correction inside the contrast's own subset (6.49) the same comparison gives $p = 0.074$, and a
sign-flip permutation over the nine pairs gives $p = 0.095$ on a mean paired difference of
$+21.1$ points. The direction is unambiguous and nine clustered pairs cannot close it, which is
what the design supports.

The roster instrument sees the same contrast on the same 360 responses at a smaller absolute
scale: 33 of 180 (18.33\%) against 9 of 180 (5.00\%), OR 4.27, Fisher $p = 1.1\times10^{-4}$,
Rao-Scott $p = 0.031$. Both instruments agree on sign and on rank; they disagree on level,
because a roster-derived rate measures overlap with a list (Section~\ref{sec:roster-instrument})
and the primary outcome measures naming.

Weaker grounding would predict fewer citations in the local language. The opposite holds:
Baltic responses average 7.55 citations in English and 12.26 in the local language.

Table~\ref{tab:lang} gives the full within-city picture under both outcome definitions.

\begin{table}[h]
\centering
\caption{Naming rate by query language within city. $p$-values are Rao-Scott at the within-arm
design effect.}
\label{tab:lang}
\begin{tabular}{llccc}
\toprule
City & Local language & English & Local & $p$ \\
\midrule
\multicolumn{5}{l}{\emph{Primary outcome}} \\
Vilnius   & Lithuanian & 47.1\% (113/240) & 19.4\% (35/180) & \textbf{0.046} \\
Tallinn   & Estonian   & 38.3\% (69/180)  & 19.4\% (35/180) & 0.162 \\
Amsterdam & Dutch      & 19.0\% (80/420)  & 1.7\% (3/180)   & \textbf{0.0007} \\
Warsaw    & Polish     & 29.0\% (122/420) & 36.7\% (66/180) & 0.224 \\
Pooled    &            & 30.5\% (384/1260)& 19.3\% (139/720)& \textbf{0.027} \\
\midrule
\multicolumn{5}{l}{\emph{Roster instrument}} \\
Vilnius   & Lithuanian & 20.8\% (50/240)  & 3.3\% (6/180)   & \textbf{0.0076} \\
Tallinn   & Estonian   & 22.8\% (41/180)  & 6.7\% (12/180)  & \textbf{0.021} \\
Amsterdam & Dutch      & 0.2\% (1/420)    & 1.1\% (2/180)   & 0.22 \\
Warsaw    & Polish     & 0.0\% (0/420)    & 0.0\% (0/180)   & undefined \\
Pooled    &            & 7.3\% (92/1260)  & 2.8\% (20/720)  & 0.067 \\
\bottomrule
\end{tabular}
\end{table}

\begin{figure}[htbp]
\centering
\includegraphics[width=\textwidth]{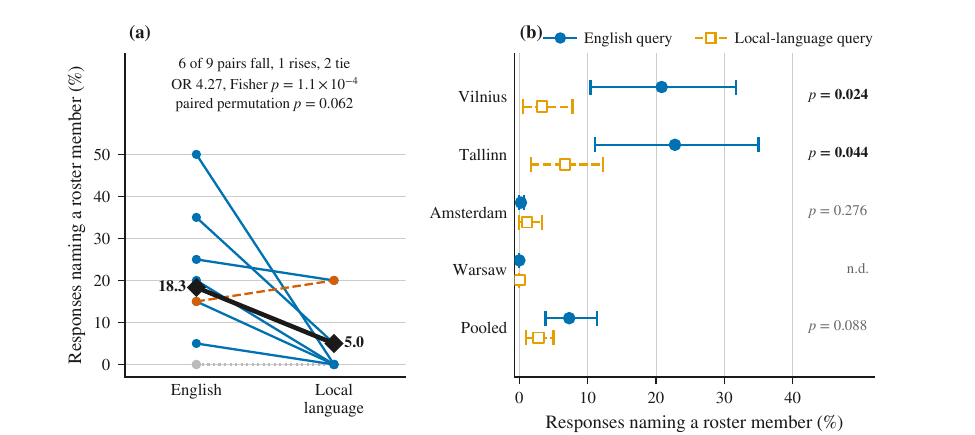}
\caption{Naming rate by query language, roster-matched outcome in both panels. Panel (a) gives
the nine matched translation pairs, 18.33\% against 5.00\%; panel (b) holds city fixed and pools
all prompts. Table~\ref{tab:lang} carries the same comparisons under the primary outcome, where
the matched pairs run 36.67\% against 15.56\%. The two instruments agree on sign and rank and
differ on level, for the reason Section~\ref{sec:roster-instrument} gives.}
\label{fig:lang}
\end{figure}

The pooled direction holds under both definitions and clears 0.05 under the primary one. Which
city carries it does not agree between the two. The roster instrument puts the effect in Vilnius
and Tallinn and cannot see Amsterdam at all, because Amsterdam has three roster matches in 600
responses. The primary outcome puts the largest effect in Amsterdam and loses Tallinn to the
Baltic design effect near eight.

Warsaw is the informative cell. Under the roster instrument it is 0 against 0, formally
undefined at any sample size. Under the primary outcome it is a well-powered null inside a
high-rate market: 29.0\% English against 36.7\% Polish, $p = 0.22$, with the point estimate
pointing the other way. A language effect that is real in Amsterdam and Vilnius and absent in
Warsaw is heterogeneous across cities, and no general property of non-English querying will
explain it. That settles the question the original analysis left open, which was whether the
language contrast could be separated from roster coverage at all.

\subsection{Naming concentration}
\label{sec:concentration}

\begin{figure}[htbp]
\centering
\includegraphics[width=0.85\textwidth]{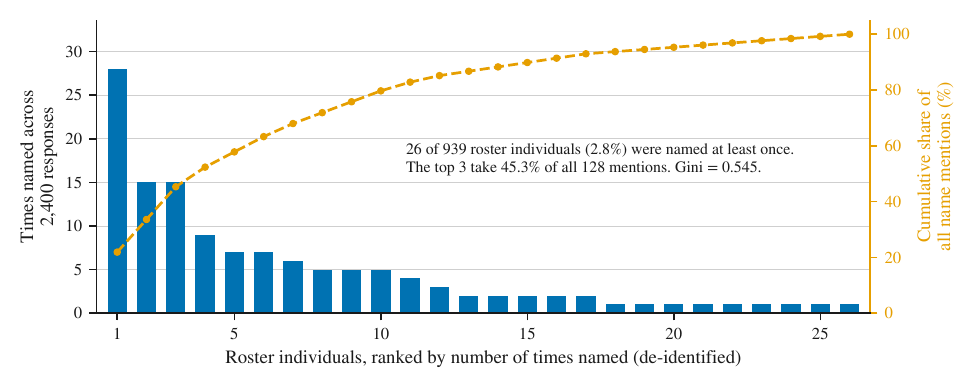}
\caption{Concentration of naming across the 26 individuals ever named.}
\label{fig:concentration}
\end{figure}

Of the 939 roster members, 26 (2.8\%) were named at least once, across 128 mentions in 114
responses. The distribution among those 26 is steep: the most-named individual takes 21.9\% of
mentions, the top three 45.3\%, the top five 57.8\% and the top ten 79.7\%, with a Gini
coefficient of 0.545 across the named and 0.987 across the full roster
(Figure~\ref{fig:concentration}). Fourteen of the 26 are named once or twice.

Prior work at brand level reports a mean Gini of 0.28 for recommendation share across fifty
brands \citep{zatuchin2026categoryownership}. Individual naming in this corpus is far more
concentrated than that. Whatever produces it does not spread across a market.

\subsection{Retrieval geography and eponymous firms}
\label{sec:geo}

\textbf{Wrong-city retrieval.} Ask a grounded model for the best car dealer in Warsaw and some of
what comes back is in Indiana. This is a retrieval failure with a measurable size, and this
section measures it.

Every generic-TLD domain carrying three or more citations was adjudicated against its live site,
its stated address, the register record where the site was unclear, and the path of the page the
model actually cited: 296 domains and 6{,}671 citations, each decided from the cited page
itself. Seventy-five domains resolve to the United States. Forty-four of them are local
businesses inside the prompted city's American homonym, and they carry 355 citations:
dealerships in Warsaw, Indiana; insurance agencies in Amsterdam, New York; car franchises in
Dublin, Ohio. A further 37 citations sit on European or global domains whose cited page is
American, so the unit that has to be adjudicated is the page
(Section~\ref{sec:geo-pages}). Together that gives 367 citations to an American local business,
1.78\% of the corpus, and 780 citations to an American page of any kind, 3.79\%. One hundred and
eighty-two of the 2{,}400 responses (7.6\%) cite at least one American local business, and 294
(12.2\%) cite an American page of some kind.

The concentration is where the mechanism shows. Table~\ref{tab:geo} gives the American local
share by city. Warsaw runs at 4.7\%, Dublin 1.9\% and Amsterdam 1.4\%, against Vilnius 0.2\% and
Tallinn 0.1\%. Vilnius and Tallinn have no American twin, and their leakage is a twentieth of
Warsaw's. Polish car dealers are the worst cell in the study at 10.3\% American local and 15.1\%
American in total, because Warsaw, Indiana holds a Toyota store, a GMC store, a Nissan store, a
Chrysler store, a Chevrolet store and several used lots, and every one of them turns up in
answers about Poland. A second mechanism operates without a namesake town: Tallinn's dealer Ideal
Auto collides in name with Ideal Auto Sales of Waukesha, Wisconsin, and an Ideal Auto in Minot,
North Dakota, and both American pages enter the Baltic cell.

\begin{table}[h]
\centering
\caption{Wrong-city retrieval by prompted city. ``American local'' counts citations whose cited
page is that of a business in the city's American homonym; ``any American page'' adds national
directories, brand locators, American job searches and other United States pages. The unit is the
cited page, so a page on a European or global host counts where the page itself is American.
Denominator is every citation issued for that city.}
\label{tab:geo}
\begin{tabular}{lrrrrr}
\toprule
City & American local & Share & Any American page & Share & Citations \\
\midrule
Warsaw    & 231 & 4.68\% & 416 & 8.43\% & 4{,}932 \\
Dublin    &  55 & 1.91\% & 182 & 6.32\% & 2{,}882 \\
Amsterdam &  70 & 1.35\% & 129 & 2.49\% & 5{,}183 \\
Vilnius   &   6 & 0.18\% &  33 & 1.02\% & 3{,}244 \\
Tallinn   &   5 & 0.12\% &  20 & 0.46\% & 4{,}339 \\
\midrule
Corpus    & 367 & 1.78\% & 780 & 3.79\% & 20{,}580 \\
\bottomrule
\end{tabular}
\end{table}

An automated first pass flagged 27 domains on a name-plus-generic-TLD heuristic. It was right on
23 of them and wrong on four, and the four wrong ones carry 153 citations, more than a third of
the flagged volume: a London insurance broker whose cited pages are all European locale paths
served by its Estonian and Polish subsidiaries, a Kraków agency on a \texttt{.com} domain, and
two Dutch brokers registered with the KvK. Running the same adjudication over the 266 domains the
heuristic never flagged found 55 more American ones worth 495 citations, more than the entire
flagged set. The heuristic was both wrong on its heaviest calls and blind to the larger problem,
which is why the adjudication is evidence-based throughout. Two domains sit outside the European
and American categories, a Vancouver broker and an American rating agency's global ranking, and
they are counted in neither. Seven domains carrying 52 citations returned low confidence, and
every one was counted as European, which is the conservative direction for a contamination
estimate.

\label{sec:geo-pages}
\textbf{The host is the wrong unit.} Adjudicating hosts misses a second route. Indeed, Yelp,
Facebook, LinkedIn, TikTok, Reddit, Tripadvisor and the App Store all serve the European market
and all of them served American pages into these answers: Indeed job searches in Warsaw, Indiana
and Dublin, California (12 citations); the Yelp category page for real-estate agents in Dublin,
Ohio (5); the Yelp listing for Toyota of Warsaw, Indiana (5); the Facebook page of Nissan of
Warsaw (2); a LinkedIn mechanic-jobs page for Warsaw, Indiana; a TikTok tag; a Reddit thread from
Indianapolis; Kia's American dealer locator; and the App Store page for an American app called
Ideal Auto, which shares its name with a Tallinn dealer. Thirty-seven citations across ten
domains, 12 of them naming one American firm. A host-level rule scores every one of them as
European, and a reader following the link lands in the United States.

At the level of the naming outcome, dropping detections that resolve to American individuals or
businesses removes 23 of Ireland's 119 naming responses, 7 of Poland's 195, 5 of the
Netherlands' 88 and 0 of the Baltics' 252. Ireland falls from 28.3\% to 22.9\%, Poland from
32.5\% to 31.3\%, the Netherlands from 14.7\% to 13.8\%, and the market ordering changes from
Poland $\approx$ Baltics $>$ Ireland $>$ Netherlands to Baltics $\approx$ Poland $>$ Ireland $>$
Netherlands. A second count works from the response text. The share of responses naming any
American state, spelled-out country or American twin city, counting responses that name nobody,
is 42.4\% in Ireland, 25.8\% in Poland, 9.2\% in the Netherlands and 1.2\% in the Baltics.

The contamination does not explain the result. Responses citing an American local business name
an individual at 31.9\% against 25.3\% for the rest, and deleting every one of those 182
responses moves the corpus naming rate from 25.79\% to 25.29\%.

One model handles the ambiguity differently. GPT-5.6 Sol declines to answer pending
disambiguation in 66 of its 105 Irish calls, 43 of 150 Polish, 31 of 150 Dutch and 0 of 195
Baltic. The other three answer, and some of them answer about Ohio.

\textbf{Eponymous firms.} Two hundred and forty-nine of 1{,}876 detections (13.3\%) are firm
names shaped like personal names. This is overwhelmingly Irish: 212 of 444 Irish detections
(47.8\%), against 6.6\% in the Netherlands, 2.2\% in Poland and 1.7\% in the Baltics. In 68 of
the 197 Irish responses carrying any detection (34.5\%), every detection is an eponymous firm.
Without the eponym rule Ireland has the highest naming rate of any market (46.9\%); with it, the
third. The same structure appears in the roster-matched data: of the five naming events outside
the Baltics, four are agency brands whose roster rows are agency owners.

Where a firm is named after its principal, ``the model named a person'' and ``the model named a
company'' are not separable propositions from the output string alone. Any individual-level
AI-visibility metric has to say which of the two it counts. This one counts the person, at a
measured cost in recall.

\subsection{RQ3: cross-model agreement, by three definitions}
\label{sec:rq3}

Thirty-one (prompt, iteration) tuples have two or more of the four models naming at least one
roster-matched individual. Among those 31, all naming models produced an identical name set in 9
cases (29.0\%, 95\% CI 16.1 to 46.6); every naming model's set shared at least one person in 14
(45.2\%); and at least one pair of naming models overlapped in 16 (51.6\%).

The word ``identical'' belongs only to the first of the three. Prior work at company level
reports 41.6\% cross-model agreement on the top-recommended brand under the same replication
design \citep{zatuchin2026categoryownership}. That figure is an identity comparison on a single
top-ranked entity, so the comparable quantity here is 29.0\%.

The comparison does not survive its own denominator. The 31 tuples come from 11 prompts and 19
individuals, and a prompt-cluster bootstrap on the 45.2\% figure returns an interval of 17.4 to
71.4, which contains the 41.6\% brand-level comparator and most other values a reader might
propose. Roughly 1{,}500 tuples would be needed to separate the two. RQ3 is reported here as a
descriptive note with its interval, and it carries no conclusion.
\citet{constantin2026consensus} measured cross-engine consensus on 278 production responses to
56 buyer-intent prompts across seven cities and five systems, and reports convergence
conditional on category with no single pooled rate, which bounds the question without settling
it.

\section{What a roster-based instrument measures}
\label{sec:roster-instrument}

Individual-level AI visibility is naturally measured against a roster: assemble a list of real
professionals, ask the model, count matches. This study ran that design alongside the
roster-free one on the identical 2{,}400 responses. The comparison is worth reporting because
the roster design produces a clean, plausible and misleading result.

Across the corpus the models emitted 27{,}293 name-shaped mentions. The 939-person roster
matched 128 of them, 0.469\%, spread across 114 responses. Against the individuals the
roster-free detector resolves, roster capture is 128 of 1{,}876, 6.8\%. Twenty-six of the 939
people were ever named. The resulting rates are 0.0\% in Poland (0 of 600, Clopper-Pearson upper
bound 0.61\%), 0.48\% in Ireland, 0.50\% in the Netherlands and 13.97\% in the Baltics, with one
cell, Baltics real estate, carrying the result at 25.38\% (66/260, cluster-adjusted interval
17.7 to 33.9). Only that cell and Baltics car dealers survive Holm correction across the twelve
cells. The 0.0\% to 25.4\% range is a roster-overlap statistic.

Read as model behaviour, the table says grounded models never name a real individual in Poland.
Read as an instrument property, it says a 255-person LinkedIn-derived sample of the Polish market
intersected the models' actual naming behaviour zero times, while the models named individuals
in 31.3\% of Polish responses. The second reading is the correct one, and the diagnostic that
separates the two is not available inside the roster design: 27{,}165 unmatched name-shaped
mentions were discarded as extraction noise, and the observation that they do not affect the
roster-matched result is exactly why that result cannot answer RQ1.

Restricting the roster to the prompted city sharpens the point. Every prompt names a city; the
roster does not, so a response can match a member who works 300 km away. Against city-filtered
rosters the Baltic rate falls from 13.97\% to 8.46\% and the Dutch rate from 0.50\% to 0.33\%.
Ireland is not testable, because its one named row carries the market-name fallback in its
location field. Within the Baltics the two cities behave differently: 78.6\% of Vilnius's named
responses attach to a roster row that mentions Vilnius, against 26.4\% for Tallinn, so Tallinn's
rate is substantially a country-level rate reported under a city label.

This caution applies to any roster built this way. A roster instrument reports the
intersection of two populations, and its measured rate falls with the size of the market being
sampled, which manufactures a market effect ordered by market size. Here the roster design
orders markets Baltics $\gg$ Netherlands $\approx$ Ireland $>$ Poland, and the roster-free design
orders them Baltics $\approx$ Poland $>$ Ireland $>$ Netherlands. A roster-based product metric
for individual AI visibility, sold on the promise that it tells a professional whether models
name them, would have reported zero to every Polish subscriber in this sample while the models
were naming Polish individuals in nearly a third of answers.

\section{Discussion}
\label{sec:discussion}

\subsection{What predicts naming}

Table~\ref{tab:glmm} gives the multivariable model. Every fixed effect in it survives the random
intercept on prompt, and the ordering matches the marginal tables.

\begin{table}[h]
\centering
\caption{Logistic mixed model, random intercept on prompt, outcome = named an individual
(primary). Odds ratios with 95\% intervals. Reference levels: Baltics, real estate, Gemini,
English. Random-effect standard deviation on prompt 1.30. A GEE with an exchangeable working
correlation clustered on prompt returns the same signs and the same significant terms, with
$\alpha = 0.243$.}
\label{tab:glmm}
\begin{tabular}{lcc}
\toprule
Term & OR & 95\% CI \\
\midrule
Market: Poland        & 1.29  & 1.06 to 1.57 \\
Market: Ireland       & 0.39  & 0.30 to 0.51 \\
Market: Netherlands   & 0.20  & 0.16 to 0.27 \\
Category: car dealers & 0.91  & 0.76 to 1.08 \\
Category: insurance   & 0.08  & 0.06 to 0.11 \\
Model: GPT-5.6 Sol    & 3.51  & 2.80 to 4.41 \\
Model: Perplexity     & 8.61  & 7.00 to 10.61 \\
Model: Grok           & 11.29 & 9.19 to 13.87 \\
Query language: local & 0.30  & 0.24 to 0.38 \\
\bottomrule
\end{tabular}
\end{table}

Category and model dominate. Insurance cuts the odds of naming by a factor of twelve against
real estate, and the model answering the question moves them by a factor of eleven from bottom
to top. Market sits below both, and query language cuts the odds by a factor of three.

The citation result changes shape once source domains are classified on what they are. Naming
responses and firm-only responses cite the same number of sources, and they cite firm-owned
pages at the same rate. The mechanism proposed in the original analysis of this corpus, that a
person becomes visible because their name is legible on the employer's domain, is not supported.
What separates the two is a surface the individual controls, plus a category portal, plus the
absence of social platforms.

The honest reading is narrower than the original and points the other way. A named individual's
visibility tracks a first-party surface the individual owns. The employer's own website predicts
nothing. The effect is small, 2.6 points of citation share on 619 naming responses, and it is
stable across four stress tests, so a stratified replication with more naming events in more
than one market is the necessary next step.

\subsection{What this says about measuring individual AI visibility}

Three properties of this corpus complicate any individual-level metric, and none of them is
specific to this study.

The first is the roster problem of Section~\ref{sec:roster-instrument}. Any instrument defined
by membership of a list reports the intersection of that list with model behaviour, and reports
it as if it were model behaviour.

The second is the eponym problem of Section~\ref{sec:geo}. In categories where firms are named
after their principals, and Irish estate agency is one, the output string does not decide
whether a person or a company was named. Forty-eight per cent of Irish person-shaped detections
are firm brands, and the rule that removes them changes which market ranks first.

The third is retrieval geography. A grounded model asked for the best car dealer in Warsaw
returns dealerships in Warsaw, Indiana, and a metric that counts names without checking where
they are will count them. Correcting for it moves Ireland by 5.5 points.

A measurement that ignores all three will still produce a number, and the number will be wrong
in a direction that varies by market.

\section{Ethics, data protection and competing interests}
\label{sec:ethics}

\textbf{Human-subjects status.} No person took part in this study. Nobody was recruited,
enrolled, interviewed or surveyed, and no intervention ran on anyone. The experiment issued
2{,}400 calls to four commercial APIs and compared the returned text against a list of publicly
listed professionals. Nobody was contacted, and nobody was harmed.

No institutional ethics committee reviewed this study. The work was designed and run by
Rankfor.AI OÜ, which is the controller, and it follows the Estonian Code of Conduct for Research
Integrity. Estonian law makes prior committee review compulsory for research that processes
special categories of personal data without consent (Personal Data Protection Act, \S\,6(4)).
This study has no participants and no special-category data, so that requirement does not apply.

\textbf{Personal-data status.} ``No human participants'' and ``no personal data'' are separate
claims. The first holds here. The second does not. Public availability settles nothing on its
own, and Zimmer's account of the Tastes, Ties and Time release is the standing demonstration
\citep{zimmer2010public}.

The roster is personal data. It holds 939 living, identifiable people in Poland, Ireland, the
Netherlands, Lithuania and Estonia, each with a name, a public professional headline, a city or
region, and a resolving \texttt{linkedin.com/in/} URL (Section~\ref{sec:method-roster}). It came
from public LinkedIn search results on 24 July 2026. It carries no email address, no telephone
number, no purchased record and no special-category data. Those 939 people are data subjects,
and building and querying that file is processing, whatever its source. The stored model output
is a second surface, since it names people outside the roster. The lawful basis is Art.~6(1)(f)
GDPR, legitimate interest.

\textbf{Legitimate interest assessment.} A written three-part assessment was completed before
publication. Purpose: whether grounded models name individual professionals is unmeasured, and
it bears on how a class of workers becomes visible to buyers
(Section~\ref{sec:related-individual}). Necessity: the study asks whether a model names
\emph{real} people, so synthetic or pseudonymised names cannot answer it. Balance: subjects and
their employers published this data so that prospective clients would find it, the processing
yields no decision about any individual, nobody was contacted, and 26 of the 939 were ever named
by any model. Against that, the subjects do not know, could not object beforehand, and the
aggregate result speaks to their professional visibility. The assessment concludes that the
interest carries, conditional on the safeguards below, and marks Art.~14 transparency as its
weakest limb. Rankfor.AI's data protection officer reviewed and signed off the assessment,
including the Art.~14(5)(b) reliance and the retention period, before submission.

\textbf{Article 14 transparency.} The data was obtained indirectly, so Art.~14 applies and its
clock runs one month from collection. The study relies on the research exemption in
Art.~14(5)(b), and the reliance is contestable: 939 is a small cohort and the data was days old,
both of which weigh against a disproportionate-effort claim (Recital~62). No contact details are
held, and acquiring 939 of them would mean collecting more data than the study needs.
Art.~14(5)(b) requires appropriate measures from any controller relying on it, and names
publication as one. A standing notice at \url{https://open.rankfor.ai/resources/research-data-protection-notice} gives the
controller, source, purpose, lawful basis, retention and objection route, published whether or
not a supervisory authority accepts the exemption.

\textbf{Safeguards, retention and objection.} Art.~89(1) safeguards apply
\citep{edpb2026research}. Four fields per person were collected and no more. The design is
pseudonymised throughout: individuals appear in the analysis and in the deposit as opaque
identifiers, and no personal data enters the public deposit. The identifying roster sits in a
private, access-controlled file under one named custodian, apart from the analysis outputs, and
it is deleted by 31 July 2028 unless a fresh assessment justifies keeping it. Objection and
erasure requests go to \texttt{dpo@rankfor.ai} and are honoured on receipt; a request removes
the individual from the roster and from any retained response text, and the affected aggregate
figures are recomputed and corrected on the record. No individual is named in this paper, its
figures or its supplementary materials.

\textbf{Competing interests.} The author holds a commercial interest in Rankfor.AI, a vendor
selling AI-visibility measurement and services, and the dual affiliation appears on the title
page. The interest is specific and points in a particular direction: an individual-visibility
gap is what such a product exists to close. Three design features bound what that interest could
do to the numbers, and they are named here so a reader can check each one. The
inclusion rule was fixed before collection. The primary outcome was moved off the roster-derived
measure whose headline supported the commercial framing most strongly, and
Section~\ref{sec:roster-instrument} sets out why. The citation claim that made the cleanest
sales argument, that a person becomes visible through the employer's own website, is reported
here as not supported. The aggregate dataset, the domain coding, the coding manual and the
detector source are published, so every figure can be recomputed. No client of Rankfor.AI
appears in the roster or the analysis.

\textbf{Funding.} Rankfor.AI paid the API and tooling costs. No external, public or client
funding supported this work, and nobody outside the author approved the manuscript before
submission.

\textbf{Author contributions (CRediT).} Dmitrij Żatuchin: conceptualization, methodology,
software, validation, formal analysis, investigation, data curation, writing (original draft),
writing (review and editing), visualization, project administration, funding acquisition.

\section{Limitations}
\label{sec:limitations}

\begin{enumerate}
    \item \textbf{Detector recall bounds every rate from below.} The detector demands positive
    evidence before accepting a candidate, which buys precision (96.9\%) at the cost of recall
    (61.7\% against the 128 roster mentions, 62.6\% in the Baltics). Every naming rate here is a
    lower bound, and the shortfall is uneven across markets: a hand audit of the dropped pool
    puts the genuine-individual share at 36.4\% in the Baltics against 4.5\% in the Netherlands,
    which compresses the market ordering reported above and does not create it.

    \item \textbf{Eponymous firms and retrieval geography are handled by rule.} 13.3\% of
    detections are firm names shaped like personal names, concentrated in Ireland (47.8\%), and
    the rule that removes them changes which market ranks first. Residual unresolvable ambiguity
    is roughly 10\% of detections, 7.2\% flagged and about 3.1\% silent, concentrated in the
    Netherlands. The geographic stage rests on adjudication of 296 domains, and a false negative
    there inflates the market it sits in. The adjudication covers generic-TLD domains with three
    or more citations; below that line sit 346 hosts carrying 933 citations, 265 of them cited
    once, and European country-code domains were not adjudicated because a \texttt{.pl},
    \texttt{.ie}, \texttt{.nl}, \texttt{.lt} or \texttt{.ee} address cannot be an American
    homonym.

    \item \textbf{Differential grounding.} 177 responses (7.38\%) returned no citation, 137 of
    them from one model and 78 of them in one market. The grounded-only sensitivity run reverses
    nothing, and the confound is real.

    \item \textbf{Single time point and exploratory replication depth.} All 2{,}400 calls were
    issued inside a two-hour window on one day, at the exploratory tier of five iterations
    \citep{zatuchin2026diceroll}. Grounded model behaviour varies with the web corpus a search
    tool indexes at query time, and this study makes no claim about temporal stability. The
    clustering correction, not the nominal $n$, is what bounds precision: effective $n$ is 407
    on the primary outcome and 396 on the roster-matched one.

    \item \textbf{The prompt sets are not translation-matched throughout.} Of seven prompt slots
    per Baltic category, three are true English/local translation pairs. The pooled
    English-against-local comparison therefore mixes a language contrast with a prompt-wording
    contrast, which is why Section~\ref{sec:language} leads on the nine matched pairs and reports
    the pooled figures as descriptive.

    \item \textbf{Convenience roster.} The roster reflects what LinkedIn's Classic-mode search
    surfaced. Section~\ref{sec:roster-instrument} treats that as the object of study, and it
    means no roster-derived rate here is a population estimate.

    \item \textbf{Cross-model agreement is indeterminate.} The 31 tuples behind the 45.2\%
    overlap figure come from 11 prompts and 19 individuals; the prompt-cluster interval runs 17.4
    to 71.4, and two alternative identity rules on the same tuples give 29.0\% and 51.6\%.

    \item \textbf{The design cannot resolve a demographic dimension.} Work on platform-mediated
    visibility reports gender disparities near an incidence rate ratio of 0.90
    \citep{hannak2017bias}. With 26 individuals ever named, the smallest gap this design could
    detect at 80\% power is roughly 25 percentage points, and an effect of the size that
    literature reports would be detected here with probability 0.06. No demographic result is
    reported, and none should be read into the silence.

    \item \textbf{Several contrasts are out of reach, and are named as such.}
    Table~\ref{tab:power} lists the comparisons this design cannot support at 80\% power. Real
    estate against car dealers, Poland against the Baltics, and Grok against Perplexity are
    nulls the design could not have resolved.
\end{enumerate}

\begin{table}[h]
\centering
\caption{Power at the primary outcome. Responses per arm needed at 80\% power and
$\alpha = 0.05$, after inflation by the within-arm design effect, against what the design
supplies.}
\label{tab:power}
\begin{tabular}{lrrl}
\toprule
Contrast & Needed per arm & Have & Verdict \\
\midrule
Grok vs Gemini                 & 87       & 600 & powered \\
Amsterdam: English vs Dutch    & 106      & 252 & powered \\
Perplexity vs Gemini           & 128      & 600 & powered \\
Real estate vs insurance       & 195      & 800 & powered \\
Car dealers vs insurance       & 229      & 800 & powered \\
Poland vs Netherlands          & 280      & 600 & powered \\
GPT-5.6 Sol vs Gemini          & 359      & 600 & powered \\
Grok vs GPT-5.6 Sol            & 362      & 600 & powered \\
Baltics vs Netherlands         & 536      & 678 & powered \\
\midrule
Perplexity vs GPT-5.6 Sol      & 697      & 600 & under \\
Vilnius: English vs Lithuanian & 378      & 206 & under \\
Tallinn: English vs Estonian   & 704      & 180 & under \\
Ireland vs Netherlands         & 1{,}260  & 494 & under \\
Warsaw: English vs Polish      & 1{,}369  & 252 & under \\
Poland vs Ireland              & 1{,}434  & 494 & under \\
Baltics vs Ireland             & 2{,}596  & 546 & under \\
Grok vs Perplexity             & 7{,}375  & 600 & under \\
Real estate vs car dealers     & 25{,}915 & 800 & under \\
Poland vs Baltics              & 215{,}737& 678 & under \\
\bottomrule
\end{tabular}
\end{table}

\section{Conclusion}
\label{sec:conclusion}

Grounded language models name an individual professional in about a quarter of answers to
buyer-intent questions, and what moves that share is not what a marketing account of it would
predict. Category moves it most: insurance cuts the odds of naming by a factor of twelve against
real estate, in every market tested. The model answering moves it almost as much, from 9.3\% for
Gemini to 38.0\% for Grok on identical prompts. Query language moves it by a factor of three,
and on nine exact translation pairs English named an individual 36.7\% of the time against
15.6\% for the same question asked in Lithuanian or Estonian. National market, the variable a commercial
reader reaches for first, comes last and separates only the Netherlands from the rest.

When a model does name a person, the citation trail that leads there runs through a surface the
individual controls and through category portals. The employer's own website carries no signal
at all. That reverses the mechanism the first analysis of this corpus proposed, and it holds
under four stress tests: a change of outcome definition, stratification, a geographic
correction, and the withdrawal of every domain whose only claim to being one person's site was
that a person's name appeared in it.

The measurement lesson has the widest reach. A roster is the obvious instrument for this
question, and it fails at it. A 939-person roster built under a defensible public-data rule
matched 0.47\% of the name-shaped text the models produced and 6.8\% of the individuals they
named, surfaced 26 people, and reported zero for an entire national market in which the models
were naming individuals in nearly a third of answers. Two further properties would distort any
naive count: half of Irish person-shaped output is agency brands named after their founders, and
a measurable share of answers about European cities describe their American homonyms. Anyone
building an individual-level AI-visibility metric has to solve all three before the number means
anything, and the shortest route to a defensible one runs through detection from the text. A list
can only find the people already on it.

\section{Data and code availability}
\label{sec:data}

Aggregated, feature-coded response-level data is published on Zenodo under CC BY 4.0 at
\url{https://doi.org/10.5281/zenodo.21612690}: market,
category, city, query language, model, iteration, citation count, source-type counts and both
naming indicators. The deposit also holds the domain coding table, the coding manual, the detector source and the
geographic adjudication record, documented as a datasheet \citep{gebru2021datasheets}. The
coding table carries 2{,}117 rows against the 2{,}182 domains analysed here: 66 person-named
hosts are masked for the reason given below and pooled under one label.

Every result in Sections~\ref{sec:results} onward recomputes from those files, with two
exceptions stated here: the per-cell roster sizes in Table~\ref{tab:roster} and the filtered
candidate pool behind the Polish stemmed re-check derive from the withheld roster and the
withheld response text, and appear in this paper only.

Response text is not published, because it contains third-party personal data the subjects did
not provide (Section~\ref{sec:ethics}). The identifying roster, names and LinkedIn profile URLs,
is not published and is held in an access-controlled research file. Individuals a model named
appear as 26 opaque pseudonyms. A re-identification audit tests the linkage route: no response
flagged as naming an individual cites a host built from that individual's own name, which
required masking 66 such hosts (508 citations, 2.5\% of the citation table) across naming and
firm-only responses alike. Masking is applied globally, so the flag is uninformative overall:
3.1\% of citations on naming responses against 2.4\% otherwise, Fisher $p = 0.20$. Within the
Baltics, where naming concentrates, a response carrying a masked host is 2.7 times more likely to
be a naming response ($p = 0.0004$). That conditional dependence is stated because every analysis
here is stratified by market. Anyone who re-runs a public prompt may recover some names
from the model itself; median per-call reproduction across the 57 (person, prompt) pairs where
naming occurred is 5\%. That residual is accepted because the alternative is withholding the
concentration and cross-model results entirely.

The README documents the environment (Node 22.19.0, \texttt{openai} 6.26.0,
\texttt{@google/genai} 1.43.0), the four API keys and the exact commands. Cost was about
USD 24 at July 2026 list prices. The pipeline records no token counts, so that figure is
list-price arithmetic on measured call volumes and no billed total is available.

\appendix

\section{Prompt set}
\label{sec:prompts}

All 120 prompts appear verbatim in the deposit, grouped by market, category and language, with
the prompt identifier used in the response-level table. The nine matched translation pairs of
Section~\ref{sec:language} are flagged there.

\section{Detection cascade}
\label{sec:pseudocode}

The detection cascade of Section~\ref{sec:method-matching} is published as source and as
pseudocode in the deposit, including the blocklists, the gazetteer, the eponym test, the
geographic stage and the fuzzy-match thresholds used by the roster matcher.

\bibliographystyle{plainnat}

\end{document}